\documentstyle[epsfig]{aipproc}

\begin{document}

\title{Morphology and Dynamics of High Z Radio Galaxies and 
Quasars
\footnote{To appear in ``After the Dark Ages: When Galaxies Were Young'', 
          Ninth Annual October Astrophysics Conference, 
          ed. S.S. Holt \& E.P. Smith, 
          American Institute of Physics Press, 1998}}
\author{K. C. Chambers}
\address{Institute for Astronomy, University of Hawaii,
         2680 Woodlawn Drive, Honolulu, HI 96822}


\maketitle

\begin{abstract}
The continuum morphologies of high redshift radio galaxies and
quasars can be modeled as enormous bipolar reflection
nebulae from shells of dust swept up by bipolar outflows.  
If the observed shape of a particular object is fit with an
analytic function, then the velocity of the shell is specified 
by the equations of motion.  
The predicted kinematics can be compared with the observed
emission line velocity field, and the resulting fit is excellent.  
The implications for massive galaxies at high redshift include 
the requirement of an initial epoch of star formation that creates 
dust distributed throughout a very large, diffuse, nearly virialized halo. 
\end{abstract}

\section*{Physical Model of the Alignment Effect}

High redshift radio galaxies have peculiar spatially extended 
optical and infrared morphologies that are generally aligned 
along the axis of their powerful radio sources
\cite{cham87,mcca87}.  
This phenomena is called the ``alignment effect'' 
\cite{cham90a}. 
Dozens of distinct mechanisms for this phenomena have been proposed in
the literature,  
but there has been no general consensus on the nature of the emission mechanisms.  
Among the various proposals for an optical alignment effect is the idea  
that the surrounding medium scatters a narrow anisotropic beam of light 
(e.g. a blazar) from an AGN  
and this beam appears as a visible pencil of light, 
like a searchlight beam scattered by fog
\cite{tadh89,fosb89,fabi89,dise89}.
Tadhunter et al.\  \cite{tadh89} argued that 
if the scattering medium had an optical depth $\tau \sim 0.1$, 
then the scattered intensity would be consistent with the brightest
known blazars. 
This idea is inextricably mixed with proposals for the unification 
of radio sources \cite{bart89}.
In the unified scheme, radio galaxies are 
the unbeamed parent population of QSRs: 
a radio galaxy observed within 
45$^\circ$ of the axis  would have the appearance of a QSR. 

Optical polarization has been detected in a number of HZRGs 
\cite{dise89,scar90,janu91}.
Spectropolarimetry has found specific QSR features 
in the polarized component of the spectra of HZRGs  
\cite{cima96,dey96,dise96}.
Furthermore, 
broad band optical and infrared polarimetry show the polarization properties are 
consistent with the characteristics of scattering by 
silicate-graphite dust grains rather than electron scattering 
\cite{knop97}.
The discovery of polarization in HZRGS in the infrared  
\cite{knop96b}
is crucial because it shows that the light redward of the 4000 \AA\ 
break can be dominated by dust scattering. 
This is an important step forward in understanding the 
infrared alignments seen in HZRGs 
\cite{cham88,eise89,cham96b}.  

Manzini \& di Serego Alighieri \cite{mdsa96} proposed a specific 
model for dust scattering in high redshift radio galaxies 
where a diffuse spherical halo of dust was illuminated
by a 45 degree bi-conical beam of quasar light, rather than
a narrow blazar beam. Either model has  
a fundamental problem if the scattering medium is dust.
They assume a unified scenario where radio galaxies are 
the parent population of the radio quasars, but 
if this assumption is true, 
these models would predict the radio quasars 
would be reddened when observed near the axis. 
This is not observed. 

A solution to the problem can be found if the dust is distributed
in an expanding bipolar shell of dust with an evacuated interior 
\cite{cham99}.  
If the outward opening shell is illuminated by the active nucleus, 
then we can recover a self-consistent unification scheme. 
Although this is an oversimplification,  
I postulate that the morphologies and polarization properties 
are due to giant bi-polar dust nebulae \cite{cham96}.  
For the high redshift radio galaxies, this hypothesis   
can account for an extraordinarily wide range of phenomena including:   
the alignment effect, the various features of quasar nebulosity, 
the presence of quasar spectral features in the spectropolarimetry of 
of high redshift radio galaxies, and the unification of quasars and
radio galaxies without requiring reddening of quasars observed on axis. 

In order to investigate this proposal further,
I have modeled the morphology and dynamics of expanding 
bipolar dust shells illuminated from a central source
\cite{cham99}.  
By assuming an analytical form for an axisymmetric 
density distribution $\rho = Q(\theta)r^{-2}$, 
and bipolar wind force $P(\theta)$, 
the shape of a swept-up shell 
can be determined by quadrature \cite{cham90b,cham99}.
If $P(\theta)$ and $Q(\theta)$ and the orientation $i$ 
are chosen such that the resulting shell matches the observed morphology, 
then the model predicts the dynamics of the shell. 
The predicted velocity field can be compared with long slit spectroscopy. 

An example of this kind of model is shown in Figure 1. 
The model has an inclination of 20 degrees from the plane of the sky and 
dust with a Henyey-Greenstein phase function. 
The functions $P(\theta)$ and $Q(\theta)$ were chosen to fit
the morphology of the high redshift radio galaxy 3C265 \cite{long94}.
The model spectrum has four components, 
two from the isotropic insitu photoionized emission line gas from the  
front and back surfaces of the shell, 
and two from the nuclear line emission scattered by dust grains 
swept up in the expanding shells. 
The model gives a remarkably good fit to the complex data set
\cite{tadh91}. 

This model is the first physically self-consistent explanation 
of both the morphology and dynamics of high redshift radio galaxies  
\cite{cham99}. 
Any alternative model for the complex emission line spectrum, 
e.g. entrainment in the jet, triggered star formation, or mergers, 
would be unlikely to reproduce these features seen in the velocity
field data without introducing excessively ad-hoc components. 
Alternative scattering mechanisms for the polarization 
such as electron scattering or inverse Compton are similarly excluded. 
A prediction of the model is that the two scattered emission line components 
will be partially polarized, whereas the two isotropic emission line
components will not be polarized.
Furthermore the scattered components should contain the high ionization
lines of the broad line region.  
The continuum emission should show the same partial polarization
as the scattered components (which will not be the same as the
total line emission since some fraction is insitu and some fraction is isotropic).

\section*{Implications for Galaxy Formation}

The strong evolutionary ``turn on'' of the alignment effect at $z\sim 1$
together with the ubiquity of the phenomena up to at least $z \sim 5$ has not
been addressed in standard cosmogonies.   
This aspect of the alignment effect is particularly noteworthy
given the success of the dynamical model discussed above. 
The model is largely dependent on the assumption of
dust distributed through out a very large halo with an ambient 
density distribution roughly proportional to $1/r^2$.  
This implies a large, nearly viralized halo diffused with dust from
a previous star formation episode. 
In particular, the halo cannot be very ``lumpy'' as one would expect
from the merging of subgalactic units, or the dynamics and morphologies 
would be far more disorganized than they are. 
The presence of aligned structures out
to redshifts $z>4$ indicates that large halos were well organized 
at the time of the first major epoch of starformation \cite{cham99}.

\begin{figure} 
\centerline{\epsfig{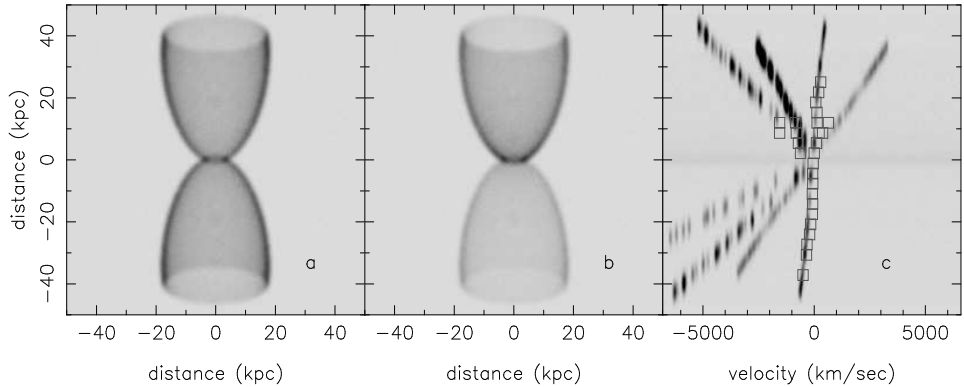}}
\vspace{10pt}
\caption{
Dynamical model of the extended emission in high redshift radio galaxies
as an expanding bipolar dust shell 
which scatters light from a quasar core (Chambers 1999). 
The functions $P(\theta)$ and $Q(\theta)$ were chosen to fit
the morphology of the high redshift radio galaxy 3C265.
A simulated narrow band image is constructed from the isotropic line
emission (a), and a simulated broad band image is constructed
from the scattered continuum (b). 
The model long slit spectrum has four components, two from the isotropic 
insitu photoionized emission line gas in the 
front and back shells, and two from the nuclear 
line emission scattered by dust grains swept up in the expanding shells. 
The predicted long slit spectrum (c) is compared with the 
observed $[O III]$ velocity field of Tadhunter (1991) shown in boxes.  
The lumpiness of the model spectrum is an artifact of the course
grid required by computing resources; the average surface 
brightness is representative. A falsifiable prediction of the model
is that the scattered spectral components 
will be partially polarized and show the high ionization line ratios of the 
BLR whereas the isotropic components will not be polarized and will have
low ionization line ratios.
}\label{myfirstfigure}
\end{figure}

\end{document}